\documentclass[12pt]{article}
\usepackage{a4}
\usepackage{epsf}
\usepackage{rotating}
\usepackage{epsfig}
\usepackage{amssymb}
\usepackage{colortbl}
\usepackage{colordvi}

\newcommand{\beq}{\begin{eqnarray}}
\newcommand{\eeq}{\end{eqnarray}}

\newcommand{\be}{\begin{equation}}
\newcommand{\ee}{\end{equation}}
\newcommand{\lwrsim}{\raise0.3ex\hbox{$<$\kern-0.75em\raise-1.1ex\hbox{$\sim$}}}

\def\C2#1#2{({\cal C}_2)_{#1}^{#2}}

\begin{document}
\setcounter{page}{1}


\title{Short comment about the lattice gluon propagator at vanishing momentum}
\author{ Ph.~Boucaud$^a$, Th. Br\"untjen$^a$, J.P.~Leroy$^a$, A.~Le~Yaouanc$^a$, A.Y.~Lokhov$^b$,\\
J. Micheli$^a$, O. P\`ene$^a$, J.~Rodr\'iguez-Quintero$^c$ and 
C.~Roiesnel$^b$ }
\maketitle
\begin{center}
$^a$Laboratoire de Physique Th\'eorique et Hautes
Energies\footnote{Unit\'e Mixte de Recherche 8627 du Centre National de 
la Recherche Scientifique}\\
{Universit\'e de Paris XI, B\^atiment 211, 91405 Orsay Cedex,
France}\\
$^b$ Centre de Physique Th\'eorique\footnote{
Unit\'e Mixte de Recherche 7644 du Centre National de 
la Recherche Scientifique\\ 
}de l'Ecole Polytechnique\\
F91128 Palaiseau cedex, France\\ 
$^c$ Dpto. F\'isica Aplicada, Fac. Ciencias Experimentales,\\
Universidad de Huelva, 21071 Huelva, Spain.
\end{center}
\begin{abstract}
We argue that all evidences point towards a finite non-vanishing 
zero momentum renormalised lattice gluon propagator in the infinite volume limit.
We argue that different simulations with different lattice setups 
end-up with fairly compatible results for the gluon propagator at zero
momentum, with different positive  slopes as a function of the inverse volume.  

\end{abstract}
\begin{flushright}
\begin{tabular}{l}
{\tt LPT Orsay 06-08}\\
{\tt CPHT-RR 009.0106 }\\
{\tt UHU-FT/06-02}\\
\end{tabular} 
\end{flushright}

\section{Introduction}

The lattice gluon propagator at small or vanishing momentum in the Landau gauge
has recently been frequently addressed  as it is related to several studies in
the small momentum regime using non-lattice methods. It is often advocated
that  the zero momentum gluon propagator should vanish, while we 
have~\cite{Boucaud:2005ce} shown a Slavnov-Taylor based argument in favor of a
divergence when the momentum goes to zero.  Notwithstanding these extraneous
arguments we observe that the genuine lattice data point towards {\it a finite
non-vanishing gluon propagator at zero momentum in the infinite volume limit}.
Our second claim is that, once a well defined renormalisation procedure has
been defined, the different available results are close enough, despite
several systematic effects, to suggest an agreement.

Our aim in this note is simply to gather the arguments  in favor of this claim,
without discussing the relationship with any non-lattice claim. We will not
present any new result but only quote published results and add a reanalysis of
our old  data. We concentrate on $SU(3)$ pure Yang-Mills theory in the Landau
gauge.

There are two approaches to the problem.
\begin{itemize}
\item One is to simply compute the gluon propagator at zero momentum and
perform a well defined renormalisation. It is well known that the result is a
finite non-vanishing value. But it might happen that the vanishing only happens
in the infinite volume limit. Therefore an extrapolation to infinite volume is
needed.
\item The second approach uses a set of small non-vanishing momenta and tries a
fit of the propagator in terms of a power law $(p^2)^{\alpha_G-1}$, or
equivalently of the dressing function in terms of $(p^2)^{\alpha_G}$. The fit gives some
range of value for $\alpha_G$. The value $\alpha_G=1$ -- which corresponds to a
non-vanishing of the gluon propagator at vanishing momentum -- has obviously
zero measure and it is thus impossible to be assertive with this second method.
It is nevertheless important to check that  $\alpha_G=1$ is compatible with the
result and to check that the gluon propagator at vanishing momentum is in
continuity with the result at small non-vanishing momenta. 
\end{itemize}

\section{Definitions and notations}

In Landau gauge the gluon propagator writes
\begin{equation}
G_{\mu\nu}(p) = (\delta_{\mu\nu}-\frac{p_\mu p_\nu}{p^2})\; G^{(2)}(p^2),
\end{equation}
which implies 
\begin{eqnarray}
G^{(2)}(p^2) &=& \frac 1 3 \sum_\mu G_{\mu\mu}(p)\quad \hbox{\rm for} \quad
p\ne 0 \nonumber\\
G^{(2)}(0) &=& \frac 1 4 \sum_\mu G_{\mu\mu}(0).
\end{eqnarray}
The factor 1/4 for zero momentum is due to an additional degree
of freedom (clearly the orthogonality to the momentum in Landau gauge does not
provide any constraint for $p_\mu = 0$) related to the fact that the Landau
gauge fixing algorithm keeps unconstrained the global gauge
transformations~\footnote{Notice that  this theoretically justified  3/4 
factor is numerically confirmed as it  ensures the continuity of the gluon
propagator at $p^2 \to 0$ which will be discussed later on.}. 

In order to be able to compare the results from  different gauge actions and
different lattice  spacings one needs to renormalise the gluon propagator. The
standard method on the lattice is the Momentum substraction scheme (MOM)  which
amounts to define the renormalised propagator 
$G_R^{(2)}$ from the bare one $G^{(2)}$ according to
\begin{equation}
G^{(2)}(p,a) = Z_3(\mu,a)\; G_R^{(2)}(p,\mu)
\end{equation}
where the renormalisation condition is 
\begin{equation}\label{MOM}
G_R^{(2)}(\mu,\mu) \equiv \frac 1 {\mu^2},\quad \hbox{\rm whence}\quad
Z_3(\mu,a)= \mu^2 \;G^{(2)}(\mu,a)
\end{equation}
where $a$ is the lattice spacing, i.e. the ultraviolet cut-off.

This renormalisation can thus be done non perturbatively from lattice data
provided that $\mu$ is in the available range for the given lattice spacing
$a$. If this is not the case it is necessary  to match the $Z_3$'s with
different lattice spacings. To illustrate this let us give 
an example: if we take $\mu=4$ GeV it is
not possible to compute directly   $Z_3$ for the Wilson gauge action with
$\beta = 6.0$ ( $a^{-1}= 1.97$ GeV)~\footnote{It is advisable to keep $p <
(\pi/2)\, a^{-1}$.}. We will thus use the results at $\beta = 6.4$ ($a^{-1}=
3.58$ GeV). We then need to compute $Z_3(\mu,6.0)/Z_3(\mu,6.4)$. This ratio is
independent of $\mu$ at leading order.  It can thus be computed non perturbatively for momenta
in which both lattice spacings provide data. An analytic approach is to rely on
the one loop perturbative formula 
\begin{equation}\label{pert}
 \frac {Z_3(\mu,a')}{Z_3(\mu,a)} = \left(
\frac{\beta(a')}{\beta(a)}\right)^{{13}/{22}}, 
 \end{equation}
which is valid for small enough lattice spacings 
(in the perturbative regime).

\section{The gluon propagator at vanishing momentum}

The Adelaide group has performed a systematic study~\cite{Bonnet:2001uh} of the
gluon propagator  in the infinite volume limit. They use the mean-field
(tadpole) improved  version of the tree-level, $O(a^2)$ Symanzik improved gauge
action. They choose a MOM renormalisation at $\mu =$ 4 GeV i.e.  $G_R^{(2)}(4
\,{\rm GeV},\mu=4 \,{\rm GeV} )= 1/(4\, {\rm GeV})^2$. They fit the volume
dependence of the zero momentum gluon propagator on several lattice spacings
and lattice volumes up to a volume of 2000 fm$^4$, with always a spatial cubic
lattice and a length in the time direction twice the spatial length. Their
fitting formula is  
\begin{equation}
G_R^{(2)}(0,\mu=4 \,{\rm GeV}) =  G_{R\,\infty}^{(2)}(0,\mu=4 \,{\rm GeV})
+ \frac c V,
\end{equation}
and gives
\begin{equation}\label{australiens}
  G_{R\,\infty}^{(2)}(0,\mu=4 \,{\rm GeV}) = 7.95\pm 0.13\, \mathrm
{GeV}^{-2},\quad c= 245\pm 22 \,  \mathrm{fm}^4\,\mathrm {GeV}^{-2}. 
\end{equation}
This result clearly indicates a finite non vanishing   $G_R^{(2)}(0,\mu)$. It is
strange that nobody objects to this published result but that, nevertheless,
one repeatedly reads that the zero momentum gluon propagator vanishes.

\begin{table}[h]
\centering
\begin{tabular}{|c|c|c|c|}
\hline  
$\beta$ &  $V$ in units of $a$    
      &bare propagator $G^{(2)}(p,a)$  & $1/L$ in GeV
\\ \hline 
5.7  &$16^4 $&$16.81\pm0.13$   & 0.0672 \\
5.7  &$24^4 $ &$15.06\pm0.29$   & 0.0448\\
5.8  &$16^4 $ &$19.12\pm0.16$   
 &0.0841\\
5.9  &$24^4 $ &$18.12\pm0.30$    &0.0685\\
 6.0  &$32^4 $ &$17.70\pm0.59$    &0.0615\\
6.0  &$24^4$  &$19.67\pm0.35$   
  &0.0821 
\\ \hline
\end{tabular}
\caption{Physical lattice sizes and raw data for the gluon propagator at zero momentum
 $G^{(2)}(p,a)$ from our old data.} 
\label{raw}
\end{table}
This is why, waiting for a systematic and extensive reanalysis, we have simply 
digged out   our old results for the gluon propagator  which have been
obtained  from simulations with the Wilson pure gauge action  on hypercubic lattices
\cite{Becirevic:1999uc, Becirevic:1999hj}. Table~\ref{raw} lists the 
normalized raw data of the gluon propagator at zero momentum for our largest
physical volumes (some of these data have never been published). No rescaling,
perturbative (Eq. \ref{pert}) or non-perturbative, has yet been applied to
these data.  Our volumes are not very  large as this was not the aim of our
simulations, and we do not claim our study to be an improvement over
ref.~\cite{Bonnet:2001uh} but simply an independent check.  Using the same renormalisation  as
ref.~\cite{Bonnet:2001uh} we find 
\begin{equation}
  G_{R\,\infty}^{(2)}(0,\mu=4 \,{\rm GeV}) = 9.1\pm 0.2\pm 0.2\, \mathrm
{GeV}^{-2},\quad c= 140\pm 30\pm 40 \; \mathrm{fm}^4\,\mathrm {GeV}^{-2}. 
\end{equation}
where the first error is statistical and the second is a systematic one
estimated from different choices of the fitting points. 

\begin{table}[h]
\centering
\begin{tabular}{|c|c|c|c|}
\hline  reference &
$G_R^{(2)}(0,\mu=4 \,{\rm GeV})$ in GeV$^{-2}$ &  $c$ in GeV$^{-2}$ fm$^4$   
      & max vol in  fm$^4$  
\\ \hline   \cite{Bonnet:2001uh} &
$7.95\pm 0.13$           &  $245\pm 22$ & 2000
\\  table \ref{raw} &
$9.1\pm 0.3$          & $140\pm 50 $   & 90
\\ \cite{Oliveira:2005hg} &
10.9 - 11.3          & 47 - 65   & 110
\\ \hline
\end{tabular}
\caption{\footnotesize Summary of the infinite volume zero momentum 
propagator and its slope in terms of $1/V$ for three different simulations.
 The largest volume used in the fit  is also indicated. The statistical error is not
 quoted in ref~\cite{Oliveira:2005hg}.}
\label{Gde0}
\end{table}
More recently ref.~\cite{Oliveira:2005hg} provides additional information on
the same issue. In their table 2 the authors report fits of the zero  momentum
gluon propagator as a function of $1/V$, using only data at $\beta=6.0$ with
Wilson gauge action obtained on very anisotropic lattices \footnote{The time length is typically 16 times the spatial one}.  The results are given in lattice units and concern bare
propagators.  

We shall assume, as has been done up to now, that the
volume dependence is polynomial in $1/V$ for large volumes. The very asymmetric shape  is meant to provide  very low values of the momentum ;   it   is interesting to  check whether the zero momentum propagator   depends on the geometry. We therefore convert the authors'  fit of the 0-momentun propagator to physical units and perform a MOM renormalisation at 4 GeV for which we use $a^{-1}( \beta=6.0)
= 1.97 \,\mathrm {GeV}$
and,  from our non-perturbative fits : 
\begin{equation}
Z_3(4 \,\mathrm {GeV}, \beta=6.0))= 1.648,
\end{equation}
and we get
\begin{eqnarray}
  G_{R\,\infty}^{(2)}(0,\mu=4 \,{\rm GeV}) &=& 11.3
  \,\mathrm{GeV}^{-2}\;\mathrm{and}\; 10.9\, \mathrm{GeV}^{-2},\nonumber
  \\\quad c= 47\, \mathrm{fm}^4\,\mathrm {GeV}^{-2} \;&\mathrm{and}&\;65\,
  \mathrm{fm}^4\,\mathrm {GeV}^{-2}.   
\end{eqnarray}
where the two results correspond to a linear/quadratic fit in
 $1/V$~\footnote{We are aware that the authors of~\cite{Oliveira:2005hg} also use
 non polynomial fits in $1/V$ which can  lead to vanishing or infinite zero
 momentum propagators. But they have themselves noticed that this destroys the
 smoothness.}.  We do not know the statistical errors.

The results of these three collaborations  are summarised in table. \ref{Gde0}.
Concerning $G_{R\,\infty}^{(2)}(0,\mu=4 \,{\rm GeV})$ the three results are in
the same ballpark and it may be conjectured that  the systematic errors are not
all taken into account: $O(a)$ effects, effect of the shape, insufficiently
large volumes (for the second and third lines),  uncertainty in the estimate of
the lattice spacing in physical units, etc.  Altogether it seems that, not only
there is a clear indication in favor of a finite non vanishing zero momentum
gluon propagator, but that different  simulations  agree on the value. Of
course a more extensive study is necessary.

Concerning the slope $c$ the numbers clearly differ, they only agree in order
of magnitude and are all positive. We expect that the slope is much more
sensitive  to systematic effects such as the shape.

 We turn now to the second approach, namely a fit of the $ p^2$ dependence of
the propagator at small momenta. We first claim that the gluon propagator is
continuous and smooth at $p=0$.  This has been observed in  several references
(see for instance figure~17 in~\cite{Bonnet:2001uh}) . This can also be seen in
figure~2  in~\cite{Tok:2005ef}. The latter paper also compares the gluon
propagator with periodic or twisted boundary conditions and concludes that the
twisted propagtor is smaller than the periodic one but that  the difference
vanishes in the large volume limit.  Let us now comment on the fit as a power
law ($G^{(2)}(p)   \propto (p^2)^{\alpha_G-1}$)  which necessarily discards the
zero-momentum.   In section 3.1 of ref.~\cite{Boucaud:2005ce} we have shown with
similar fitting formulae that $\alpha_G$ is compatible with 1 on the examples 
of $SU(2)$ and $SU(3)$. But we have experienced  instabilities  and we do not
know of any convincing results obtained with this method \footnote{We do not
understand the fits of \cite{Silva:2005hb}  and we obviously disagree with
their conclusion that the zero momentum propagator vanishes.}.  This
instability may be due to the fact that, if such a power law applies in the
small momentum limit, it can only be isolated at very small momenta which have
not yet been reached.

  \section{Conclusions}
  
  The renormalised gluon propagator at zero momentum converges, in the infinite
   volume limit,  towards a non vanishing finite value~\cite{Bonnet:2001uh} if
    one uses a volume
   dependence  which is polynomial in $1/V$~ independently of the 
   boundary conditions~\cite{Tok:2005ef}.
    We have shown that different studies with
   different gauge actions,  different parameters and different shapes of the
   volume agree rather well  on the value of the renormalised gluon propagator
   while the slopes in $1/V$ agree only in sign and order of magnitude.

     No discontinuity of the gluon propagator is seen when the momentum goes to
     zero. It results that the infrared exponent $\alpha_G$ (often named
     $2\kappa$) must be  equal to 1. We have stressed the instability of the
      fits of the propagator without the point  at $p=0$ 
      assuming a power law dependence: different fitting functions, which
      are equivalent in the infrared limit, give incompatible results. If this is
       duly taken into account in the systematic errors  the
        value $\alpha_G=1$ lies within the error bars in agreement with the
       claim about a non vanishing finite gluon propagator at zero momentum.

\end{document}